# Atomic Precision Processing of New Materials for Frontier Microelectronic Applications in High Performance Computing and Artificial Intelligence

Yevgeny Raitses,[1] Sebastian Engelmann,[2] Shahid Rauf,[3] Igor Kaganovich,[1] Jonathan Menard,[1] and Steven Cowely[1]

[1]Princeton Plasma Physics Laboratory, Princeton, NJ
[2]IBM, T.J. Watson Research Center, Yorktown Heights, NY
[3]Applied Materials, Santa Clara, CA

## I. Introduction

Current workloads on high performance computing machines are becoming more and more resource intensive. Even though 90% of Big Data created over the last 10 years has never been analyzed, it is predicted that unstructured data (main drivers: social media, Internet of Things (IOT) sensors everywhere) is growing at rates from 60-80%. These incredible growth rates in data, coupled with artificial intelligence (AI) and machine learning (ML), will lead to major increase in the market size for memory and computing power [1,2]. Many approaches are being pursued to enhance computing capability. For example, it is becoming clear that a limiting factor in scaling current architectures is the von Neumann bottleneck, where the information needs to be shuffled between logic and memory. Novel computing architectures are being explored to overcome these issues and continue to reduce the energy consumption per Gigaflop [3]. In addition, complete new approaches to fundamental computing elements are being considered ranging from memory compute approaches to advancing packaging solutions and further co-integration of multiple subsystems such as power electronics, analog devices, photonics, memory and logic circuits.

New materials can boost computation speed and efficiency in all architectures [1]. Even though new materials introduce challenges with concomitant financial risk, significant research is devoted to discovery of novel materials that can significantly boost high performance computing (HPC) and AI based workloads. Some of the most prominent examples are phase change and magnetic materials for memory and, carbon and binary or ternary semiconductors for logic [1,4]. The common theme to all of these is the necessity to introduce novel (non-Si based) materials to enable new functionalities. All these new materials bring along new processing challenges. For example, one of the biggest issues with phase change materials is the ternary nature of this material, owing to the fact that precise stoichiometries unlock novel material properties [4].

Plasma processing is one of the major fundamental technologies used for microelectronics fabrication. These new materials pose a significant challenge to plasma processing, as certain reactions are typically more favorable for some elements compared to others, leading to modifications of the original stoichiometry and in worst case also performance degradation.

Novel process technologies and approaches need to be developed to enable damage free processing of a wide range of materials. Focusing on plasma processing, active fields of research, such as atomic layer processing [5,6] (etching, deposition, material modification, cleaning) may offer a path to enabling this, however more innovation is needed to meet the ultimate goal of maintaining, tailoring and enhancing materials performance at atomic scale. Most importantly, the underlying physical and chemical processes leading to observed results are poorly understood, owing to the lack of in-situ insight in the processes occurring at the plasma-surface interface and the absence of the validated predictive modeling capabilities of these complex processes.

Finally, let us also mention that the leadership in the plasma processing for microelectronics is critically important for the US economy and national security especially in time when the US semiconductor industry is getting edged by the industry in China, Korea, and Taiwan [7].

Here, we respond to this DOE SC Request for Information (RFI) addressing point by point the questions posed in this request.

## II. Appropriateness of Topical Areas and Scope

The DOE Office of Science (SC) listed plasma science and technology as one of the basic research areas for this microelectronics initiative. We are fully supporting to have this important area in the DOE initiative on microelectronics. *The appropriateness of the plasma science/technology area for this initiative is evident from the fact that low-temperature plasmas are extensively utilized for materials processing in the modern microelectronics industry and mostly likely will continue to be the backbone of microchip manufacturing in the future.* From about 500-900 steps in manufacturing of modern microelectronic chips, about a half of these steps are done using plasma-based technologies such as etching and deposition. These plasma-based technologies are a critical enabler of the rapid advancement witnessed by microelectronics in recent decades, as evident in computers, mobile devices, displays. Future advances in microelectronics dictated by emerging and frontiers applications such as high performance computing, artificial intelligence (AI) and big data will require new materials and microchip architectures to implement most advanced microelectronic devices. In its turn, this requirement creates a new paradigm for plasma processing science that would have to provide high selectivity in complex multi-component plasma environments and atomic precision for substrate/wafer etching and deposition. There is however no readily available plasma technology which could address all these needs, but also no practical alternatives to plasma processing with respect to industrial scale production of microelectronic chips.

*Investment in the areas of plasma processing will provide U.S. leadership in almost all technology growth areas identified in this DOE Request for Information (RFI) including, but not limited to Memory and Reconfigurable Systems, Machine Learning and Artificial Intelligence, Power Electronics.*

For *Memory Systems*, the future is in most advanced resistive memory technologies. IBM is for example developing the game–changing Phase-Change Memory (PCM) technology. The key advantage of PCM over other types of memory technologies such as Random Access Memory is that it offers a multi-level storage capability. The use of ternary composite materials based on chalcogenides (e.g. $Ge_2Sb_2Te_5$ [4]), allows instead of the normal two phases—fully amorphous and fully crystalline—additional distinct intermediate states representing different degrees of partial crystallization, allowing for twice as many bits to be stored in the same physical area. The use of such an advanced memory technology would significantly boost the speed of devices and memory storage – all important for high performance computing and machine learning algorithms using large datasets.

One of the biggest challenges for the use of PCM devices is that their production at a large industrial scale with consistent quality and long endurance is not available. This is in a great part due to tremendous difficulties in maintaining the right stoichiometry between three and more chemical species need for the PCM, both in the processing plasma (or other methods) and at the wafer surface. In this respect, one of the most important fundamental questions needed to be addressed is

- *What is the effect of plasma stoichiometry on the stoichiometry at the surface and how do both affects the resulting thin (atomic scale) film properties?*

Processing of these complex new materials in plasmas requires unprecedented control over important plasma properties. The leading-edge plasma processing systems being developed by Applied Materials for Si based technologies enable better control over species fluxes, the energy distribution functions of charged species and ion to neutral flux ratio. Such control is necessary to enable atomic layer-by-layer processing capability [5,6]. These technologies include Atomic Layer Etching (ALE) [5], Atomic Layer Deposition (ALD) [6], plasma chemical etching and atomic precision cleans. However, the same degree of control for three species required for PCM has never been addressed for plasma reactors and especially for the industrial scale production. There is no clear path of how to achieve such great degree of control of plasma and plasma-surface interactions. Therefore, a key question is

- *What are the methods of control of stoichiometry in the plasma and at the surface?*

Towards answering both questions and helping in the development of enabling plasma technologies, there is a need in fundamental research of interactions between low temperature plasmas with new microelectronics relevant materials with complex chemical composition (e.g., those relevant to PCM and advanced logic devices). The thin (nanoscale) film processes on the surface should be also a part of this research. The research should integrate experimental and modeling efforts. The emphasis of experimental research should be on in situ diagnostics of the plasma, plasma-surface interactions and processes at the surface. The modeling efforts should involve multi-scale simulations from plasma to surface.

In view of all of the above, we propose DOE-SC to include the following topics in their consideration for future solicitation:

- *The development of a real-time monitoring and in situ diagnostic techniques that can provide information on plasma, substrate surface, and interaction between both during atomic precision processing of complex materials for the most advanced microelectronic devices with applications to high performance computing and artificial intelligence.*
- *The development of experimentally validated modeling tools to predict processing dynamics including plasma, chemical and material processes involved.*

**III. Effective Collaboration, Partnerships, and R&D Performers**

What is needed is a joint government-private partnership to develop the techniques required to make most advanced microchips for HPC, AI etc. Fundamental science aspects of this initiative can, in principle, be addressed by academic institutions (e.g. universities, national labs). However, the partnership with microelectronic industry and semiconductor equipment manufacturers promise to be the most effective in providing practical solutions based on fundamental science. *Therefore, the joint partnership involving national labs, universities and industry should be strongly encouraged by this Initiative.* The universities are well suited to conduct exploratory research of scientific problems and to train students. The national labs are well suited to use the most sophisticated diagnostic and computation resources for comprehensive studies of the plasmas, plasma-surface interactions and thin film growth that are the basis of the manufacturing techniques. The industry are the best suited to formulate processing challenges, equipment expertise, technology transfer and implementation of innovative ideas developed at the universities and national labs for microelectronics.

In order to encourage the university and national lab researchers to be actively working in this area, it would be important to setup intellectual property (IP) requirements which would have minimum implications and restrictions on publishing scientific results. There should be also a mechanism of sharing of IPs.

**IV. National Impact and Unique DOE Role and Contribution**

The DOE national labs are very well suited to use the most sophisticated diagnostic and computation resources for comprehensive studies of processing and generated materials. In the case of the plasma processing of semiconductor materials, a joint FES-BES program would be most suitable to address the grand challenges. Without plasma science associated with the plasma processing including in situ diagnostics and modeling of plasma and plasma-surface interactions, the production of new materials for most advanced microelectronic devices will have to rely on Edisonian, trial and error approach which is common in materials science laboratories. This will likely slow down the progress in the field and, as a result in frontiers topics of microelectronics for HPC and big data applications.

Even if applying Edisonian, trial and error approach, a number of variable parameters for modern plasma reactors are huge. Therefore, having a better knowledge of plasma science associated with the materials processing can potentially enable a substantial reduction of the search in this parameter space and thereby to speed up finding the technological solutions. In this regard, in magnetic fusion, sophisticated three dimensional codes have been developed to predict plasma parameters, and there is now a push for modeling of the whole fusion devices [8-10]. In contrast, no similar computational tools yet exist for plasma processing of semiconductor materials that would correctly account for kinetic effects in realistic 3-D space, plasma-surface interaction and gas chemistry. With the development of modern high performance computers such code development is achievable today. Comprehensive numerical tools should be experimentally validated and supplemented by a number of reduced models that allow for fast virtual prototyping of plasma processing tools.

The FES has recently established the Princeton Collaborative Research Center (PCRF) on Low Temperature Plasma led by the Princeton Plasma Physics Laboratory (PPPL). The overarching goal of this facility is to provide a unique combination of research capabilities, including most advanced plasma and surface diagnostics, computation codes, and expertise for advancing understanding and knowledge of low temperature plasma science for the academic, industrial, and national labs scientific community, with a focus on plasma-liquid and plasma-solid interactions and their control, self-organization, and synergy with other related sciences and applications. Future solicitations in Microelectronics could leverage from unique PCRF capabilities such as active spectroscopy, ns-fs time range laser scattering diagnostics – all suitable for in situ diagnostics of processing plasmas and thin films. In addition, the PCRF has a large selection of state-of-the art plasma computational tools covering a broad range of spatial and temporal scales relevant to processing applications. These computational tools include combination of fluid and particle-in-cell codes as well a variety of quantum chemistry and molecular dynamics tools needed to understand gas chemistry and plasma-surface interactions. These research resources concentrated at one of the DOE FES facility are unique and not readily available for BES national labs, individual researchers at universities and industry. It is also important for the research in this field that the PCRF has a direct access to the material evaluation facilities at the Princeton University Research Institute of Materials Science (PRISM).

## V. Program Planning and Evaluation

Developing future novel materials and processes creates equally a great opportunity as well as a great risk. One key element to minimize the risk is to evaluate all new materials by a standardized metric which can serve as the "computing performance baseline". Based on the maturity of the identified materials sets and processes, different performance metrics should be targets.

Similarly, an oversight committee both on project execution and scientific judgement would be very helpful. Alignment on project execution ensures solutions are ready at the appropriate time, while a scientific review of presented results ensures repeatability and transferability of the proposed solutions.

Lastly, frequent team update meetings help the effectiveness of funded projects. In particular annual face to face meetings (or even higher frequency) including but not limited to meetings of principal investigators of different teams, including experts in interdisciplinary fields such as plasma processing, plasma physics, chemistry and materials science will help to shape a clear picture of the current research focus needs and foster collaboration between the teams and experts involved.

## VI. Other

As much the research area is unknown at this point, many possible obstacles could be faced. Possibly the biggest impediment to impact is the large timeline and financial commitment that typically comes with capital investment. Such large investments, (e.g. processing equipment, diagnostics) usually need lots of resources and planning to meet the desired needs. An economic relief on some of the cost related to these investments could boost local economy, especially as the burden of entry is lowered.

In addition, as we mostly deal with novel materials and chemistries, a frequent exchange on safety or facilities handling related questions on these new chemicals could result in a significant improvement of quality and timeline for implementation. The safety-related learning could be offered to US-based institutions in order to provide further stimulus for economic growth.

## VII. Summary

- The plasma processing science and technology are important topics for this DOE-SC initiative on Microelectronics.

- New materials (e.g. ternary materials for memory systems) are critically important for enabling of high performance computing and AI applications. The grand challenge for plasma processing of these new materials is how to preserve useful material properties (e.g., how to keep the stoichiometry of ternary materials) when processed in plasmas.

- To address this challenge, there is a need of fundamental understanding of plasmas, plasma - surface interactions and processed film properties under relevant chemical composition of the plasma.

- The research in this area will need integrated experimental and modeling efforts of the processing. For the success of these efforts, *the Initiative should seek for the* *development of 1) in situ diagnostics of both plasma which modifies the surface, and thin films generated by such modifications, and 2) multiscale computational tools.*

- The DOE national labs are uniquely suited for a joint public-private partnership with industry to address near term and long-term challenges in microelectronics.